\documentclass{aa}  
\usepackage{hyperref}
\usepackage{scrextend}
\deffootnote[0.45em]{0.2em}{0.2em}{\textsuperscript{\thefootnotemark}\,}
\usepackage{amsmath,url}
\usepackage{algorithm}
\usepackage{amsfonts}
\usepackage{mathrsfs}
\usepackage{bm}
\usepackage{rotating}
\usepackage{color}
\usepackage{graphicx}
\usepackage{subfigure}
\usepackage{lineno}
\usepackage[toc,page]{appendix}
\usepackage[english]{babel}
\usepackage{graphicx}
\usepackage{txfonts}
\begin{document}

   \title{Signature of systemic rotation in 21 Galactic Globular Clusters from APOGEE-2}

   %\subtitle{I. Overviewing the $\kappa$-mechanism}

   \author{Ilaria Petralia\inst{1}
          \and
          Dante Minniti\inst{1,2,3}
          \and
          Jos\'e G. Fern\'andez-Trincado\inst{4}
          \and
          Richard R. Lane\inst{5}
          \and
          Ricardo P. Schiavon\inst{6}
          %\and
          %C. Ptolemy\inst{2}\fnmsep\thanks{Just to show the usage
          %of the elements in the author field}
          }

   \institute{Instituto de Astrofísica, Facultad de Ciencias Exactas, Universidad Andrés Bello, Fernández Concha 700, Las Condes,
   Santiago, Chile\\
   \email{ilariapetralia28@gmail.com}
   \and
   Vatican Observatory, Vatican City State, V-00120, Italy
   \and
   Departamento de Fisica, Universidade Federal de Santa Catarina, Trinidade 88040-900, Florianopolis, Brazil
   \and
   Instituto de Astronom\'ia, Universidad Cat\'olica del Norte, Av. Angamos 0610, Antofagasta, Chile
   \and
   Centro de Investigación en Astronomía, Universidad Bernardo O’Higgins, Avenida Viel 1497, Santiago, Chile
   \and
   Astrophysics Research Institute, Liverpool John Moores University, 146 Brownlow Hill, Liverpool L3 5RF, UK}

              %\email{wuchterl@amok.ast.univie.ac.at}
        % \and
         %   \email{c.ptolemy@hipparch.uheaven.space}
          %   \thanks{The university of heaven temporarily does not
           %          accept e-mails}

  % \date{Received September 15, 1996; accepted March 16, 1997}

% \abstract{}{}{}{}{} 
% 5 {} token are mandatory
% \abstract{}{}{}{}{} 

  \abstract
  % context heading (optional)
  % {} leave it empty if necessary  
   {Traditionally, Globular Clusters (GCs) have been assumed to be quasi-relaxed nonrotating systems, characterized by spherical symmetry and orbital isotropy. However, in recent years, a growing set of observational evidence is unveiling an unexpected dynamical complexity in Galactic GCs. Indeed, kinematic studies show that a measurable amount of internal rotation is present in many present-day GCs.}
  % aims heading (mandatory)
   {The objective of this work is to analyse {the APOGEE-2 Value-Added Catalogs (VACs) DR17 data of a sample of 21 GCs to extend the sample showing signatures of systemic rotation, in order to better understand the kinematic properties of GCs in general.} 
   Also, we aim to identify the fastest rotating GC from the sample of objects with suitable measurements.}
  % methods heading (mandatory)
   {From the sample of 23 GCs included in this work, the presence of systemic rotation was detected in 21 of the GCs, using three different methods. 
   All these methods use the radial velocity referred to the cluster systemic velocity ($\widetilde{V_{r}}$). Using the first method, it was possible to visually verify the clear-cut signature of systemic rotation. Whereas, using the second and third methods, it was possible to determine the amplitude of the rotation curve ($A_{rot}$) and the position angle (PA) of the rotation axis.}
  % results heading (mandatory)
   {This study shows that 21 GCs have a signature of systemic rotation. For these clusters, the rotation amplitude and the position angle of the rotation axis ($PA_{0}$) have been calculated.
   The clusters cover a remarkable range of rotational amplitudes, from 0.77 km/s to 13.85 km/s. }
  % conclusions heading (optional), leave it empty if necessary 
   {}

   
   %\authorrunning
   \maketitle
%
%-------------------------------------------------------------------

\section{Introduction}
Galactic globular clusters (GCs) represent one of the ideal laboratories to study stellar dynamics and its effects on
stellar evolution. Traditionally, GCs have been assumed to be
quasi-relaxed nonrotating systems, characterized by spherical
symmetry and orbital isotropy. Spherical isotropic (\citealt{King+1966}) and non-rotating (\citealt{Wilson+1975}) models have indeed been shown to provide a satisfactory zeroth-order description of the main observed dynamical properties. \\
However, in recent years, a growing set of observational evidence is unveiling an unexpected dynamical complexity in Galactic GCs, demonstrating that the traditional assumptions of sphericity, pressure isotropy and nonrotation are far too simplistic. In fact, kinematic studies show that a significant amount of internal rotation is present in many present-day Milky Way (MW) GCs (\citealt{Lane+11}; \citealt{Bellazzini+12}; \citealt{Bianchini+13};
\citealt{Fabricius+14}; \citealt{Kacharov+14}; \citealt{Kimmig+15};
\citealt{Lardo15}; \citealt{Bellini_2017}; \citealt{Boberg_2017}; \citealt{Jeffreson17}; \citealt{Cordero17}; \citealt{Lee_2017}; \citealt{Ferraro_2018}; \citealt{Kamann+18}; \citealt{Lanzoni_2018}; \citealt{Bianchini18}; \citealt{Szigeti+21}). All these results suggest that, 
since it is possible to study individual stellar components individually, the vast majority of GCs show signatures of the
presence of internal rotation. \\
This observational evidence plays a fundamental role to construct a complete dynamical picture of GCs. Indeed, the presence of a significant amount of internal rotation raises a series of fundamental issues connected to the formation and evolution of GCs and the different
dynamical processes involved (e.g., \citealt{Giersz+11}).
For instance, many studies indicate that rotation accelerates evolution (e.g. \citealt{Einsel1999}; \citealt{Kim08}; \citealt{Hong13}) and strongly shapes their present-day morphology (\citealt{van_den_Bergh_2008}; \citealt{Bianchini+13}). Moreover, the present-day signatures could be the relic of a stronger internal rotation set at the epoch of the cluster’s formation (\citealt{Vesperini14}; \citealt{Lee16}; \citealt{Mapelli17}; \citealt{Tiongco17}) or indicate a peculiar evolutionary environment (\citealt{Lane+10}, \citealt{Vesperini14}; \citealt*{Tiongco18}). Signatures of internal rotation could also be crucial in the kinematical differences between multiple stellar populations (\citealt{Richer_2013}; \citealt{Bellini_2015}; \citealt{Cordero17}; \citealt{Bellini_2018}; \citealt{Cordoni_2020a}, \citeyear{Cordoni_2020b}; \citealt{Dalessandro21}; \citealt{Libralato_2023}).
As a result of these interesting studies, new theoretical models are being developed. Of great interest is the proposed new distribution-function-based models of rotating and anisotropic models (see, e.g., \citealt{Varri&Bertin12}; \citealt{Gieles15}; \citealt{DeVita16}), expanding the traditional model. Several numerical simulations have also been carried out to understand the rotational properties of GCs (see, e.g. \citealt{Einsel1999}; \citealt{Ernst2007}; \citealt{Livernois21}, \citeyear{Livernois22}, \citealt{kamlah22}, \citealt{Tiongco22}). One of these was the starting point to find the first observational evidence of a relation between globular clusters’ internal rotation and stellar masses (\citealt{Scalco23}). As shown from these theoretical studies, the presence of internal rotation has several implications for the long-term
dynamical evolution of clusters (e.g. the observed GC rotation may be a lower limit to a GC's initial rotation due to angular momentum loss during its evolution).\\

\noindent In this work, the kinematic analysis for 23 Galactic GCs is presented, in order to verify the presence of a systemic rotation in these objects. This study is based on the individual radial velocities of stars available for each individual cluster. The data used come from the high-resolution spectroscopic survey Apache Point Observatory Galactic Evolution Experiment (APOGEE; \citealt{Maje+apogee2}). The astrometric parameters, such as position (RA and DEC, see \citealt{VB+21}), the abundances and the radial velocity referred to the cluster systemic velocity ($\widetilde{V_{r}}$) were previously calculated for all stars (see {\citealt{Schiavon23}}). For this reason, the available catalogue represents a good sample to study.\\
\noindent The present work is structured as follows. In Section \ref{2} we present the data available for the sample GCs, that were employed in the calculations. Section \ref{3} includes a kinematics analysis, while in Section \ref{4} the results are summarized and discussed, and conclusions are drawn in Section \ref{5}.

%--------------------------------------------------------------------

 \begin{table*}
\centering
\vspace{.1cm}
\caption{Summary of GCs analysed in this work. The target names and their coordinates of GC centre from 
\citealt{VB+21} catalogue are listed in columns 1-4. {The Jacobi radius for each cluster is listed in the fifth column. The number of stars available for the various clusters is shown in the sixth column, while the seventh column contains the adopted radial velocity dispersion. 
The eighth and ninth columns show the rotation amplitude values calculated using the best-fit sine function method and the peak-to-peak method respectively. The tenth column shows the values of the PA of the rotation axis,
following the methodology used to calculate $A_{fit}$. These values were
converted and reported into
PA 90 = East, anti-clockwise system. Finally, in the last column, the values of V/$\sigma$ (or $ A_{fit}/\sigma_{rv}$) calculated for each cluster. The last two GCs are clusters that do not show signature of systemic rotation.}}

  \begin{tabular}{ccccccccccc}
 
       \hline
       \hline
    Object  & & $RA_{0} $ & $DEC_{0}$  & $r_{J}$ & $N_{*}$ & $\sigma_{rv}$& $A_{fit}$  & $A_{peak-peak} $ & $PA_{0}$ & {$A_{fit}/\sigma_{rv}$ }\\ 
    &&${[\mathrm{deg}]}$&${[\mathrm{deg}]}$&${[\mathrm{deg}]}$&&${{[\mathrm{km/s}]}}$&${[\mathrm{km/s}]}$&${[\mathrm{km/s}]}$&[deg] \\
    \hline

     NGC 0104&47 Tuc&6.02379&-72.08131&{1.557}&302&12.0&9.16 $\pm$ 0.50 &12.41 $\pm${0.39}& {216} & {0.76}\\

     NGC 0362 &&15.80942&-70.84878&{0.598}& 70&8.6&3.57 $\pm$ 0.08&3.63 $\pm${0.88}&{234}& {0.42}\\

      NGC 1851 &&78.52816&-40.04655&{0.611}&71 &11.0& 5.59 $\pm$ 0.13&5.94 $\pm${5.84}&{180} & {0.51}\\

      NGC 2808 &&138.01291&-64.86349&{0.944}&132&14.1&8.42 $\pm$ 0.34 &8.29 $\pm${1.00}&{198} & {0.60}\\

       NGC 3201 &&154.40343& -46.41248&{0.925}&217&4.7&3.37 $\pm$ 0.10 &3.88 $\pm${0.26} &{324}& {0.72}\\
    
      NGC 5139 &$\omega$ Cen& 201.69699 & -47.47947 &{2.142}&1864&17.7& 13.85 $\pm$ 0.25 &12.98 $\pm${0.29} & {162}&{0.78}\\

      NGC 5272 &M 3&205.54842&28.37728&{0.714}&299&7.8&2.46 $\pm$ 0.12 &2.85 $\pm${0.40}& {162}&{0.32}\\

      NGC 5904 &M 5&229.63841&2.08103&{0.607}&259&7.8&8.03 $\pm$ 0.14&7.88 $\pm${0.74} &{198}&{1.03}\\

      NGC 6121 &M 4&245.86974&-26.52575&{1.658}&224&4.8&2.70 $\pm$ 0.08 &3.16 $\pm${0.15}&{234}&{0.56}\\

      NGC 6171 &&248.13275&-13.05378&{0.368}&65&4.1&0.77 $\pm$ 0.09 &1.38 $\pm${0.34}&{234}&{0.19}\\

      NGC 6205 &&250.42181&36.45986&{1.036} &152&9.6&5.62 $\pm$ 0.17 &6.03 $\pm${0.72}& {306}&{0.59}\\

      NGC 6254 &M 10&254.28772&-4.10031&{0.611}&87 &6.3& 2.27 $\pm$ 0.16 &3.74 $\pm${0.46}&{198}&{0.36}\\
      
     NGC 6273&&255.65749&-26.26797&{0.266}&81&12.1&2.37 $\pm$ 0.22&3.78 $\pm${2.32}&{270}&{0.20}\\
     
     NGC 6341&&259.28076&43.13594&{0.808}&80&8.7&4.18 $\pm$ 0.11 &5.31 $\pm${0.91}&{216}&{0.48}\\ 

     NGC 6388&&264.07178&-44.7355 &{0.516}&75&17.4&6.64 $\pm$ 0.47&11.68 $\pm${2.58}&{270}&{0.38}\\
      
     NGC 6397 &&265.17538&-53.67434&{1.174}& 187&5.5&1.39 $\pm$ 0.08&1.93 $\pm${0.64}&{198}&{0.25}\\

     NGC 6656 &M 22&279.09976&-23.90475&{1.308}& 412&8.9&5.75 $\pm$ 0.06 &6.24 $\pm${0.98}&{252}&{0.65} \\
     
     NGC 6715&M 54&283.76385&-30.47986&{0.618}&1809&19.2&6.01 $\pm$ 0.18 &8.20 $\pm${0.57}&{252} &{0.31}\\
    
     NGC 6752 &&287.7171&-59.98455&{0.913}& 152&7.7&1.04 $\pm$ 0.15&1.19 $\pm${0.34}&{162} &{0.14}\\

     NGC 6809&&294.99878&-30.96475&{0.549}&98&4.9&3.58 $\pm$ 0.10&3.35 $\pm${0.28}&{252}&{0.73}\\

    NGC 7078 &M 15&322.49304&12.167&{0.757}&155&13.0&4.16 $\pm$ 0.10&4.83 $\pm${1.45}&{234}&{0.32} \\
    \hline
    { NGC 6218 }&&251.80907&-1.94853&{0.508}&107&4.8&-&-&-&-\\
     {NGC 6838}&&298.44373&18.77919&{0.619}&129&2.7&-&-&-&-\\
       \hline
     \label{dati}
\end{tabular}

\vspace{.2cm}
\end{table*}

\section{Available data}\label{2}

The data used to conduct this analysis were collected from two different catalogues. The first one is a  list of globular cluster (GC) star members from the latest data release of the SDSS-IV/APOGEE-2 survey  (DR17\footnote{\url{https://www.sdss4.org/dr17}}, \citealt{Majewski_2017}; \citealt{Apogeedr17}). Cluster membership is based on a combination of position, proper motion, radial velocity, and metallicity cuts obtained using data from both the APOGEE-2 (DR17) and Gaia (EDR3) surveys \citep{Gaia}. Whereas, the second one contains the global parameters adopted for globular clusters included in the cluster member Value-Added Catalogs (VACs) generated from APOGEE-2 DR17 ({\citealt{Schiavon23}}).\\
The list of GCs and their characteristics considered in this work is shown in Table \ref{dati}.
This table lists the GC positions, number of stars, velocity dispersion, fitted and peak-to-peak amplitudes, and position angles. {The number of stars observed varies from cluster to cluster, but each cluster has at least 60 stars available. In general, the number of observed stars varies from 65 to 1864. We emphasize that the number of available stars for each cluster doubles the previous APOGEE sample.}
Further details about these APOGEE-2 data and their acquisition and reduction techniques and the criteria adopted for
selecting candidate GC members {are presented in \citealt{Schiavon23}.}

\section{Kinematic analysis}\label{3}
Among the various available parameters, of particular importance for this analysis is the radial velocity referred to the cluster systemic velocity ($\widetilde{V_{r}}$), namely the difference between the radial velocity of star ($V_{r}$) and the systemic velocity of the GC ($V_{sys}$) in units of the adopted radial velocity dispersion ($\sigma_{rv}$)\footnote{In this work, the adopted radial velocity dispersion values are referred to global dispersion. } (see equation \ref{prima}).

\begin{equation}
   \label{prima}
    \widetilde{V_{r}} = (V_{r} - V_{sys}) 
    \hspace{10mm} \mathrm{[{km/s}]}\\
\end{equation}

\noindent Indeed, all methods used in this work to check the presence of rotation in GCs are based on $\widetilde{V_{r}}$.\\

\begin{figure*}[ht]
    \centering
    \includegraphics[trim={0cm 1cm 0 0cm},clip,width=0.75\textwidth]{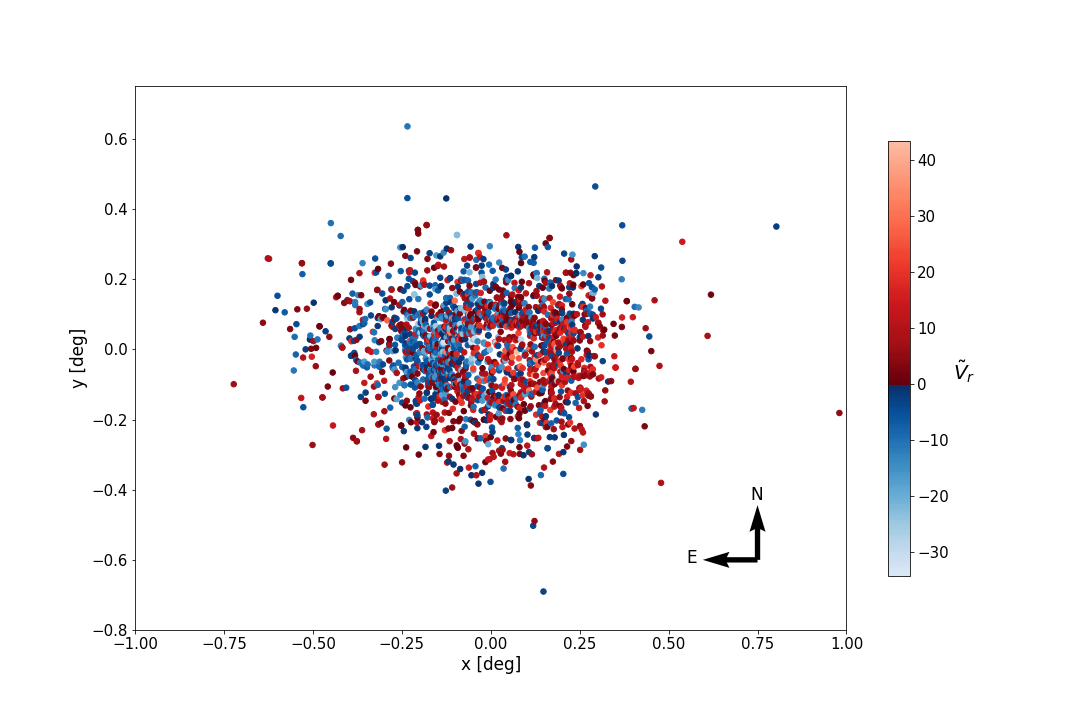}
    \caption{Distribution of the observed sample on the plane of the sky of NGC 5139, with x = ({$RA_{0}$ - RA}) $\cdot$ cos(DEC) and y = DEC - $DEC_{0}$ ($RA_{0}$ and $DEC_{0}$ being the coordinates of the cluster center, adopted from \citealt{VB+21} catalogue). The colors distinguish stars with radial velocities referred to the cluster systemic velocity ($\widetilde{V_{r}}$) > 0 (in red), from those with $\widetilde{V_{r}}$ < 0 (in blue). {Black arrow vectors indicate the north (N) and east (E) direction.}}
    \label{fig:ngc5139}
\end{figure*}

\begin{figure*}[ht]
    \centering
    \includegraphics[width=0.8\textwidth]{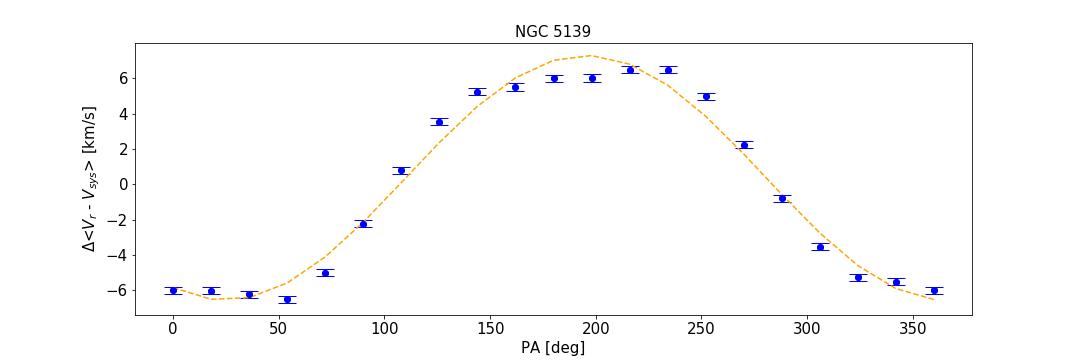}
    \caption{Difference between the mean radial velocities located on each side of the cluster with respect to a line passing through the center as a function of PA (measured from north, PA = 0°, to east, PA = 90°, clockwise system) as a function of the adopted PA, for the globular cluster NGC 5139. The orange dashed line is the sine function that best fits the observed patterns (blue dots).}
    \label{fig:ngc5139ampiezza}
\end{figure*}

\begin{figure*}

  \centering
  \includegraphics[width=.49\textwidth]{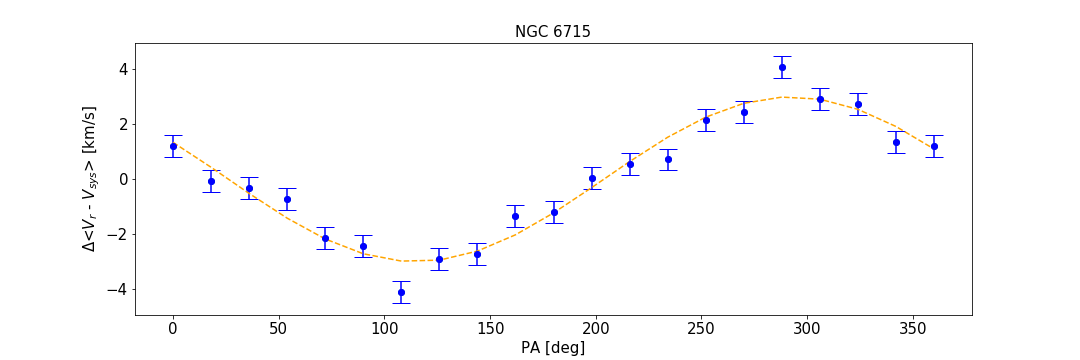}
  \hspace{0.01cm}
  \includegraphics[width=.49\textwidth]{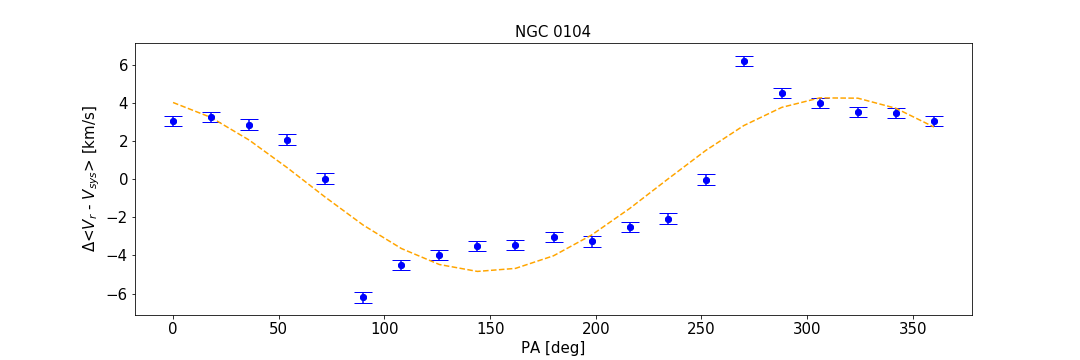}\\
   \vspace{1cm}
  \includegraphics[width=.49\textwidth]{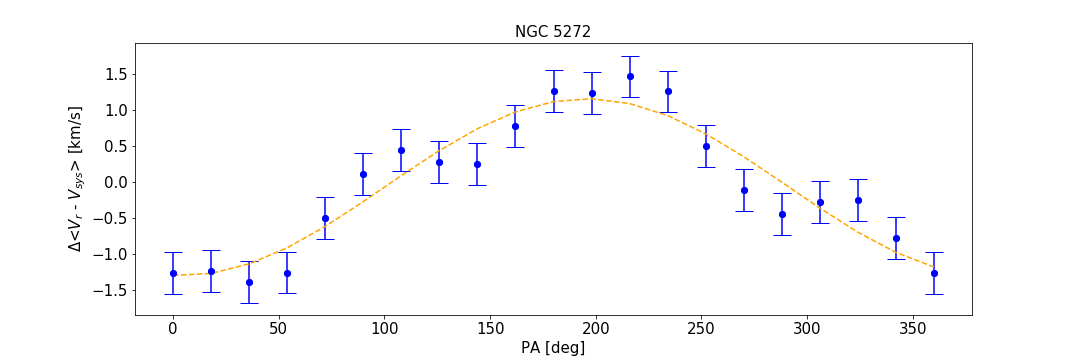}
  \hspace{0.01cm}
  \includegraphics[width=.49\textwidth]{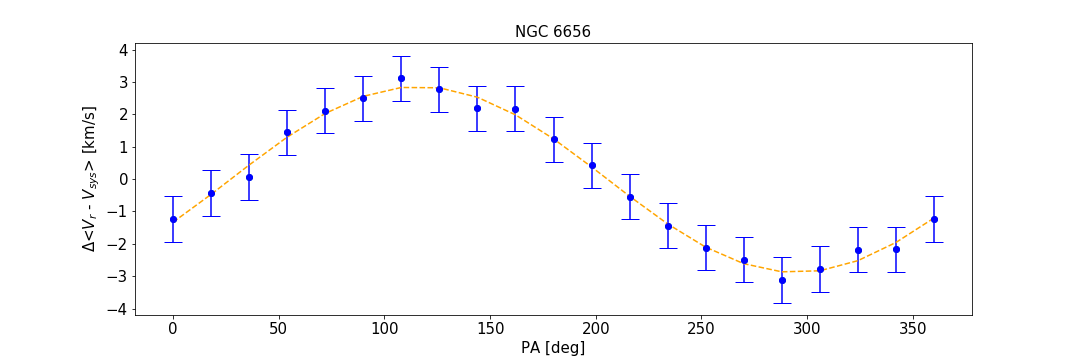}\\
  \vspace{1cm}
  \includegraphics[width=.49\textwidth]{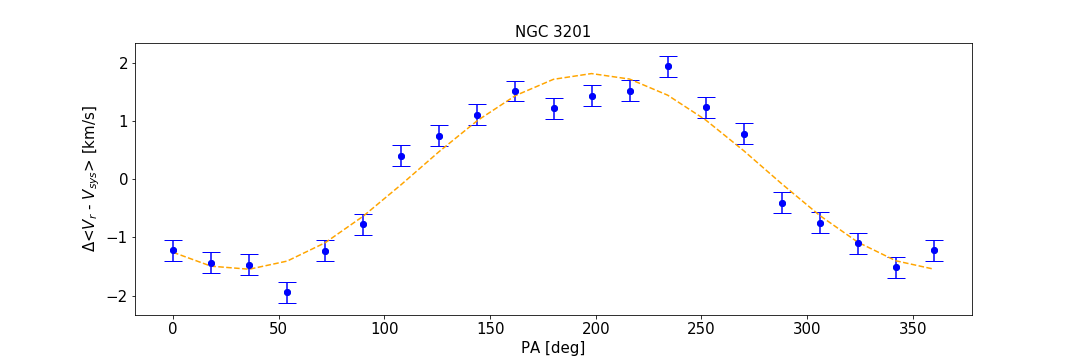}
  \hspace{0.01cm}
  \includegraphics[width=.49\textwidth]{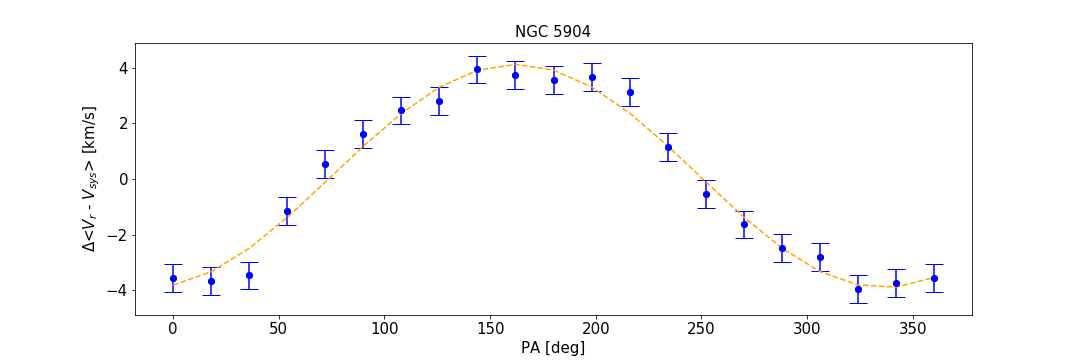}\\
  \vspace{1cm}
  \includegraphics[width=.49\textwidth]{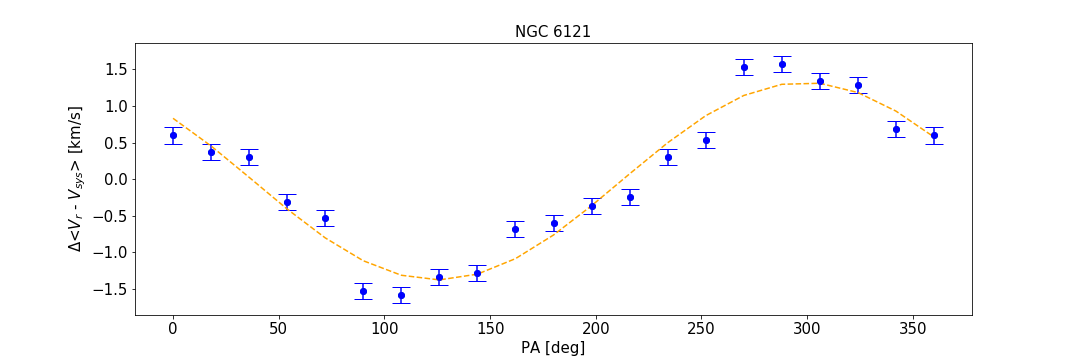}
  \hspace{0.01cm}
  \includegraphics[width=.49\textwidth]{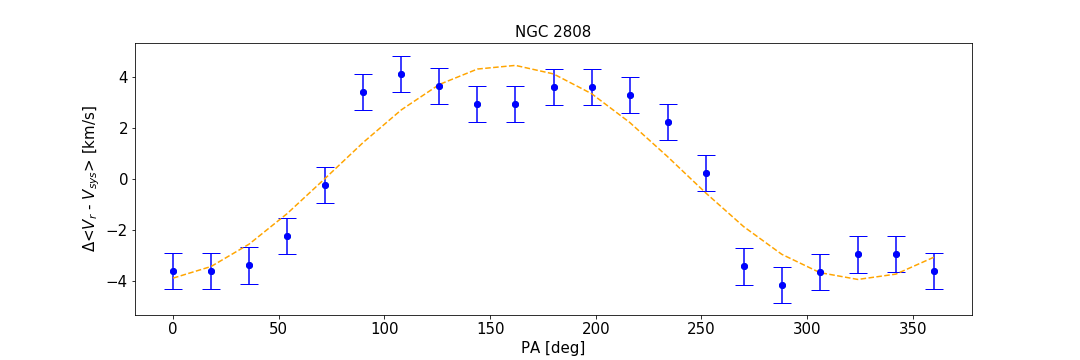}\\
  \vspace{1cm}
  \includegraphics[width=.49\textwidth]{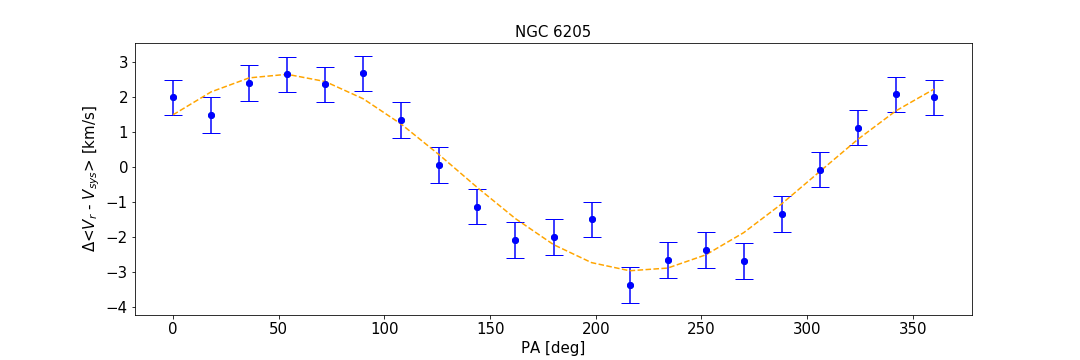}
  \hspace{0.01cm}
  \includegraphics[width=.49\textwidth]{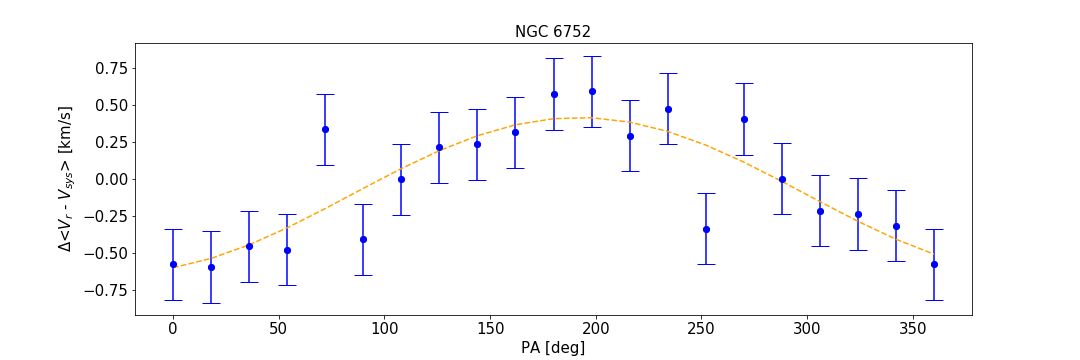}\\
  \vspace{1cm}
  \includegraphics[width=.49\textwidth]{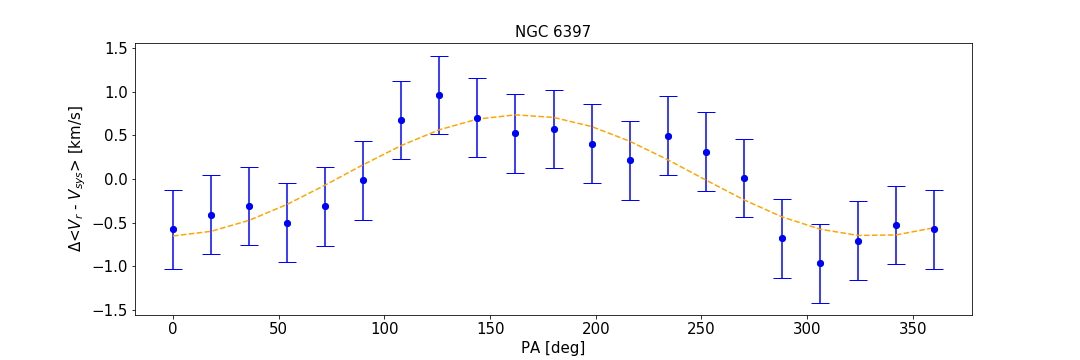}
  \hspace{0.01cm}
  \includegraphics[width=.49\textwidth]{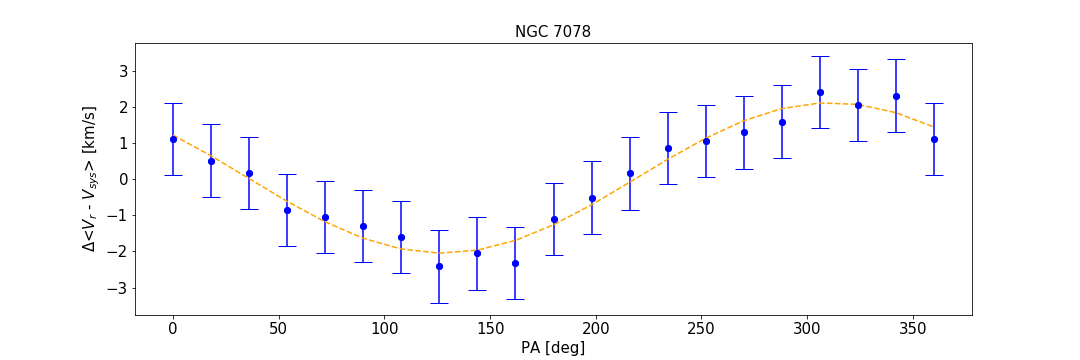}\\
  \vspace{0.5cm}
  \caption{Signature of systemic rotation for globular clusters NGC 6715, NGC 0104, NGC 5272, NGC 6656, NGC 3201, NGC 5904, NGC 6121, NGC 2808, NGC 6205, NGC 6752, NGC 6397 and NGC 7078.}
  \label{fig:rotation1}
\end{figure*}

\begin{figure*}

  \centering
  \includegraphics[width=.49\textwidth]{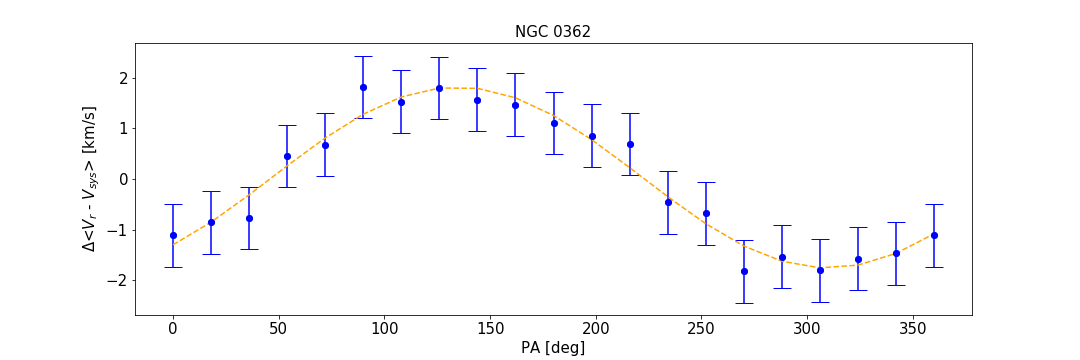}
  \hspace{0.01cm}
  \includegraphics[width=.49\textwidth]{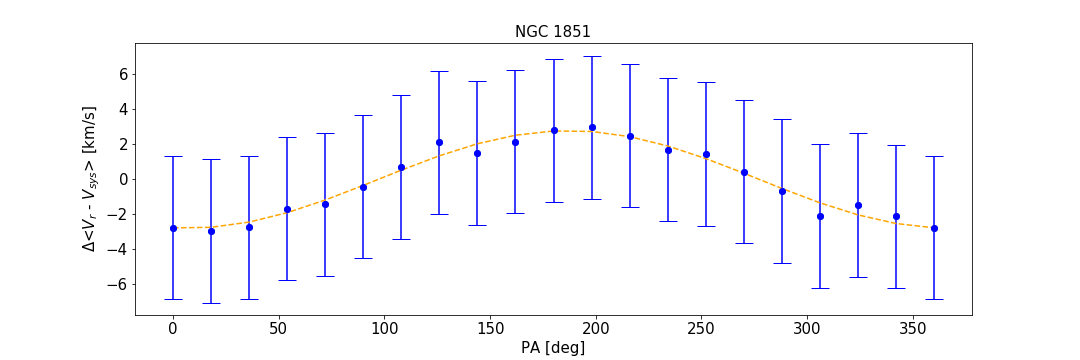}\\
   \vspace{1cm}
  \includegraphics[width=.49\textwidth]{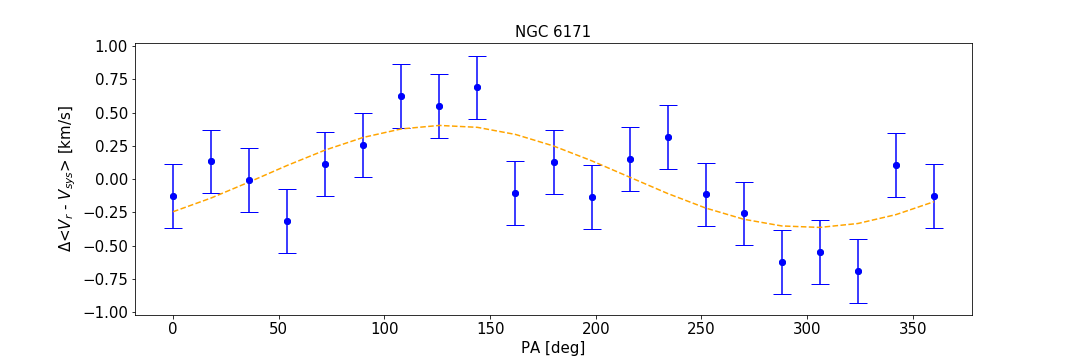}
  \hspace{0.01cm}
  \includegraphics[width=.49\textwidth]{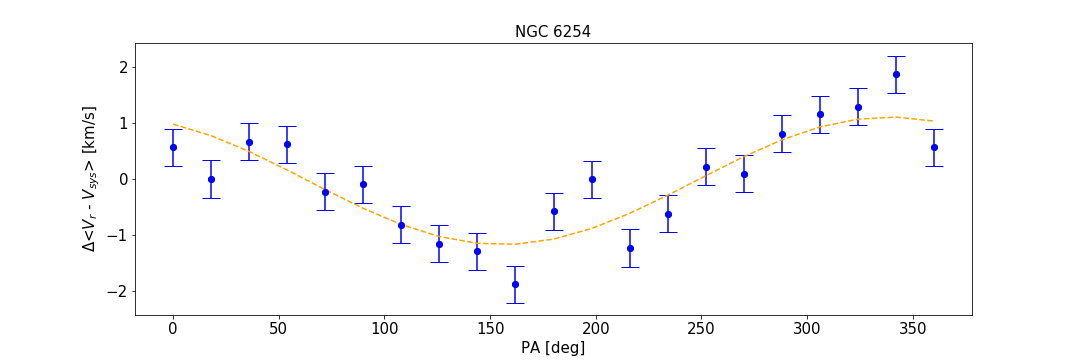}\\
  \vspace{1cm}
  \includegraphics[width=.49\textwidth]{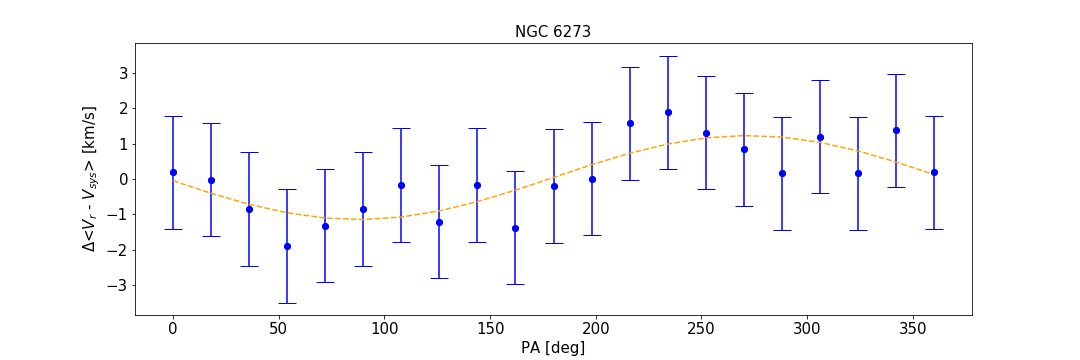}
  \hspace{0.01cm}
  \includegraphics[width=.49\textwidth]{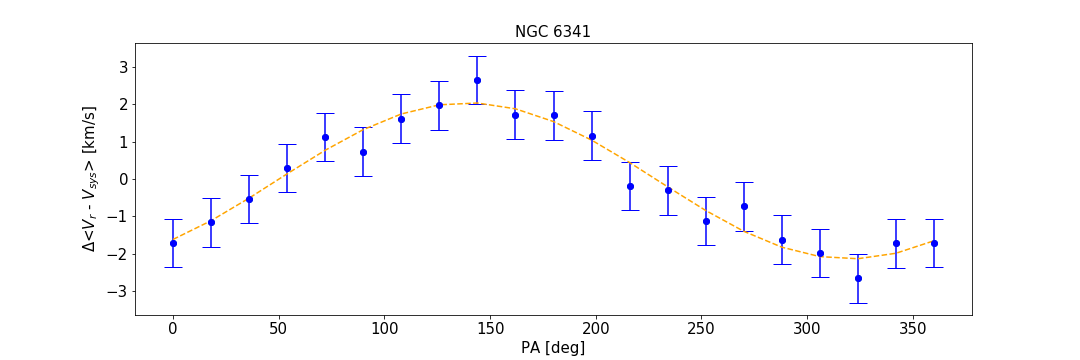}\\
  \vspace{1cm}
  \includegraphics[width=.49\textwidth]{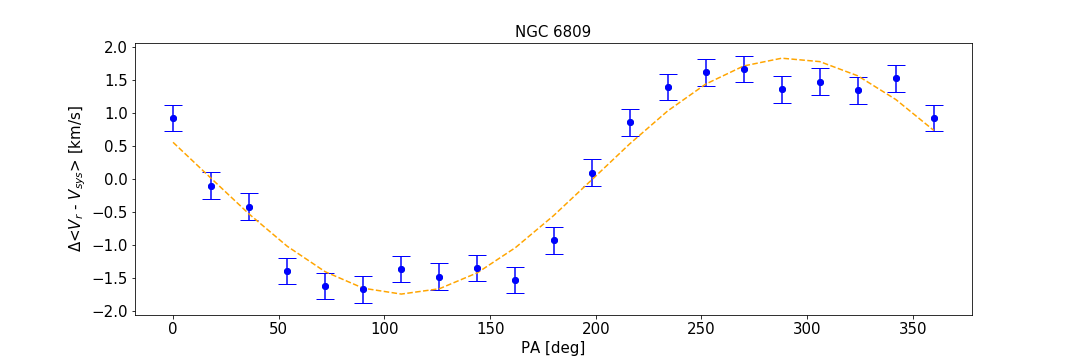}
  \hspace{0.01cm}
  \includegraphics[width=.49\textwidth]{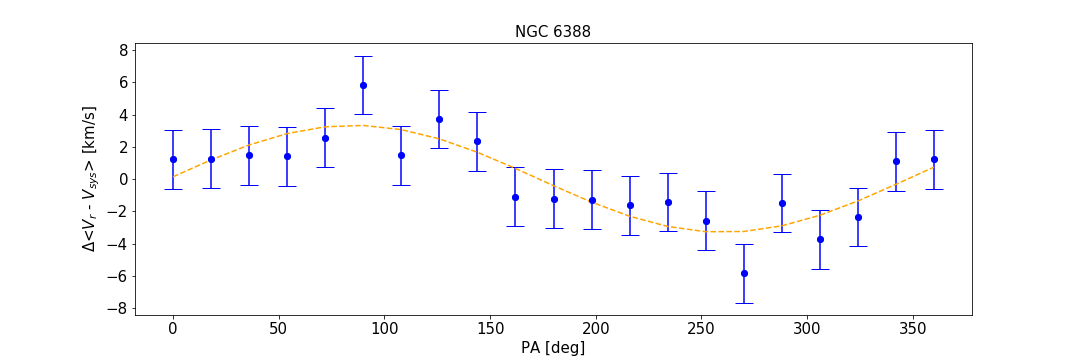}\\
  
  \vspace{0.5cm}
  \caption{Signature of systemic rotation for globular clusters NGC 0362, NGC 1851, NGC 6171, NGC 6254, NGC 6273, NGC 6341, NGC 6809 and NGC 6388.}
  \label{fig:rotation2}
\end{figure*}

\noindent A quick and visually efficient method to check for systemic rotation in GCs is to use the distribution of the stars 
with different $V_{r}$ on the plane of the sky (Figure \ref{fig:ngc5139}), with the red and the blue colors indicating, respectively, positive and negative values of $\widetilde{V_{r}}$ (i.e., $V_{r}$ larger and smaller than the systemic velocity, respectively). As apparent from the ﬁgure, the evident prevalence of stars with positive values of $\widetilde{V_{r}}$ in the lower-left portion of the map and that of sources with $\widetilde{V_{r}}$ < 0 in the upper-right part of the diagram is a clear-cut signature of systemic rotation. \\
To investigate the rotation in GCs, we used the same method, as was adopted by \citealt{Cote}, \citealt{Pacino07} and \citealt{Lane10b} and fully described in \citealt{Bellazzini+12} and \citealt{Lanzoni13}.
For any given cluster, the method consists 
in splitting the sample in two, by considering a line passing through the cluster center with position angle (PA) varying between $0^{\circ}$ (north direction) and $90^{\circ}$ (east direction), using steps of $18^{\circ}$\footnote{This value is the same for all GCs studied and was chosen to divide the range of PA values equally and while at the same time considering a significant number of stars for each range in order to perform the analysis described. {More detail explaining the choice of using this step value is shown in Appendix \ref{app_test}.}}, clockwise system\footnote{{This convention is different from that used in reference works that adopt the same analysis methodology. Thus, to compare our results with those of literature works, we convert and report the values into the reference notation, namely PA 90 = East, anti-clockwise system.}}. For each value of the PA, the difference $\Delta$<$\widetilde{V_{r}}$> between the mean radial velocity of the two sub-samples was computed and 
it is plotted as a function of the PA (see, for example, Figure \ref{fig:ngc5139ampiezza}). Its coherent sinusoidal behavior is a signature of rotation and the parameters of the best-fit sine function provide us with  the amplitude of the rotation curve and the PA of the rotation axis (for the exact meaning of this parameter see the discussion in \citealt{Bellazzini+12}). The observed patterns were fitted with the sine law, which provides a reasonable fit to the data, and is shown in equation \ref{funzioneseno}:

\begin{equation}
    \label{funzioneseno}
       \Delta<\widetilde{V_{r}}> = dist + A_{rot} \cdot sin(\omega \cdot PA  + \phi) 
\end{equation}

\noindent where $\textit{dist}$ is the displacement, $\omega$ is frequency, $\phi$ is the phase of the best-fit sine function\footnote{The values of $\textit{dist}$, $\omega$, $\phi$ have no relevance for the kinematic analysis in this work, but these coefficients are only relevant in the computational part. Indeed, they were included in the formula only to improve and find the best-fit sine function.}. $A_{rot}$ is the rotation amplitude, which is two times the actual mean amplitude, because it is the difference of the two hemispheres (in km/s). The minimum and maximum position in the best-fit sine function is the position angle of the dividing line corresponding to the maximum rotation amplitude (in degrees), coinciding with the rotation axis. 
It should be noted, however, that the values of the PA of the rotational axis ($PA_{0}$) given in the Table \ref{dati} are indicative estimates of the actual values.
The rotation amplitude derived from the best-fit function for all GCs ($A_{fit}$) is shown in Table \ref{dati}, while the observed patterns with the best-fit sine function of all GCs are shown in Figure \ref{fig:rotation1} and \ref{fig:rotation2}.\\
In this work, another method was also used to calculate the rotation amplitude, namely the peak-to-peak approach. In this case, the values of absolute minimum and absolute maximum of $\Delta<\widetilde{V_{r}}>$ among all the observed patterns are identified and the amplitude of the rotation is the difference between the maximum value  and the minimum value, namely $A_{rot} = \Delta<\widetilde{V_{r}}>_{max} - 
 \Delta<\widetilde{V_{r}}>_{min}$. The rotation amplitude obtained from this method for all GCs ($A_{peak-peak}$) is shown in Table \ref{dati}.

\subsection{{Measurement of uncertainties}}
\noindent {%Below we discuss the measurement uncertainties and their implication in assessing the rotation signals. 
As regards the best-fit sine function method, we derive the rotation amplitude and its uncertainty using the \texttt{SCIPY}\footnote{\url{https://docs.scipy.org/doc/scipy/reference/generated/scipy.optimize.curve_fit.html}} package of PYTHON. Indeed, this package allows us to derive the parameters of the best-fit sine function, such as the rotation amplitude, and their uncertainties. Whereas, in the case of the peak-to-peak approach,
the measurement of uncertainties of the rotation amplitudes was calculated starting from the uncertainties of the radial velocities of each star member of the cluster and the uncertainty of the systemic velocity of the cluster and considering the propagation of these uncertainties. The final value of uncertainty of rotation amplitude is given by the quadrature sum of the uncertainties of the observed patterns. As can be noted from the results reported in the Table \ref{dati}, the values of the uncertainty calculated for the peak-to-peak approach are slightly high, especially for some clusters with a low number of available stars.\\
In general, the APOGEE velocities are good to < 1 km/s, thus excluding the individual stars with large radial velocity errors does not show a large difference with the results reported. }\\

%\begin{multicol}

%\end{multicol}

%\begin{multicol}

%\end{multicol}

\section{Results and Discussion}\label{4}

In this work, 23 GCs were studied in order to reveal a possible presence of systemic rotation. In this analysis,  the rotation patterns were detected from radial velocities.
Note that the rotation amplitudes detected from $V_{r}$ samples are just lower limits to the true 3D amplitude, because of projection onto the plane of the sky \citep{Bellazzini+12}. Moreover, it must also be noted that the amplitude generally varies significantly with distance from the cluster center \citep[see, for example, the case of $\omega$ Cen][]{Sollima09}, but the $A_{rot}$, derived using the full range of radius, is a reasonable proxy for the actual maximum amplitude (see \citealt{Pacino07}).\\
As a main result, we find that 21 GCs show a systemic rotation and cover a remarkable range of rotational amplitudes, from $\sim$ 0.77 km/s  to $\sim$ 13.85 km/s. {For these clusters, we have also derived the PA of the rotation axis. The $A_{rot}$ and $PA_{0}$ values of these clusters are shown in Table \ref{dati}. } \\

\noindent {Once the clusters that show the signature of systemic rotation have been detected and their $A_{rot}$ and $PA_{0}$ have been calculated, we have identified the fastest rotating GCs in our sample. As reported in \citealt{Bellazzini+12}, \citealt{Kacharov+14} and \citealt{Alfaro-Cuello_2020}, the parameter $A_{rot}$/$\sigma$ (or with the most commonly used nomenclature V/$\sigma$), namely the rotation amplitude divided by velocity dispersion of the cluster, can be used for this purpose. Indeed, this parameter is expected to capture the relevance of rotation with respect to random motions. In this work, the parameter $A_{rot}$/$\sigma$ was calculated using the rotation amplitude derived from the best-fit sine function and the adopted radial velocity dispersion ($A_{fit}$ and $\sigma_{rv}$ in the Table \ref{dati}). In Table \ref{dati}, the values of this parameter are shown for each cluster of the sample. According to the results obtained, the fastest GCs are NGC 5904, NGC 5139 and NGC 0104.}\\

\begin{figure*}
    \centering
    \includegraphics[width=1\textwidth]{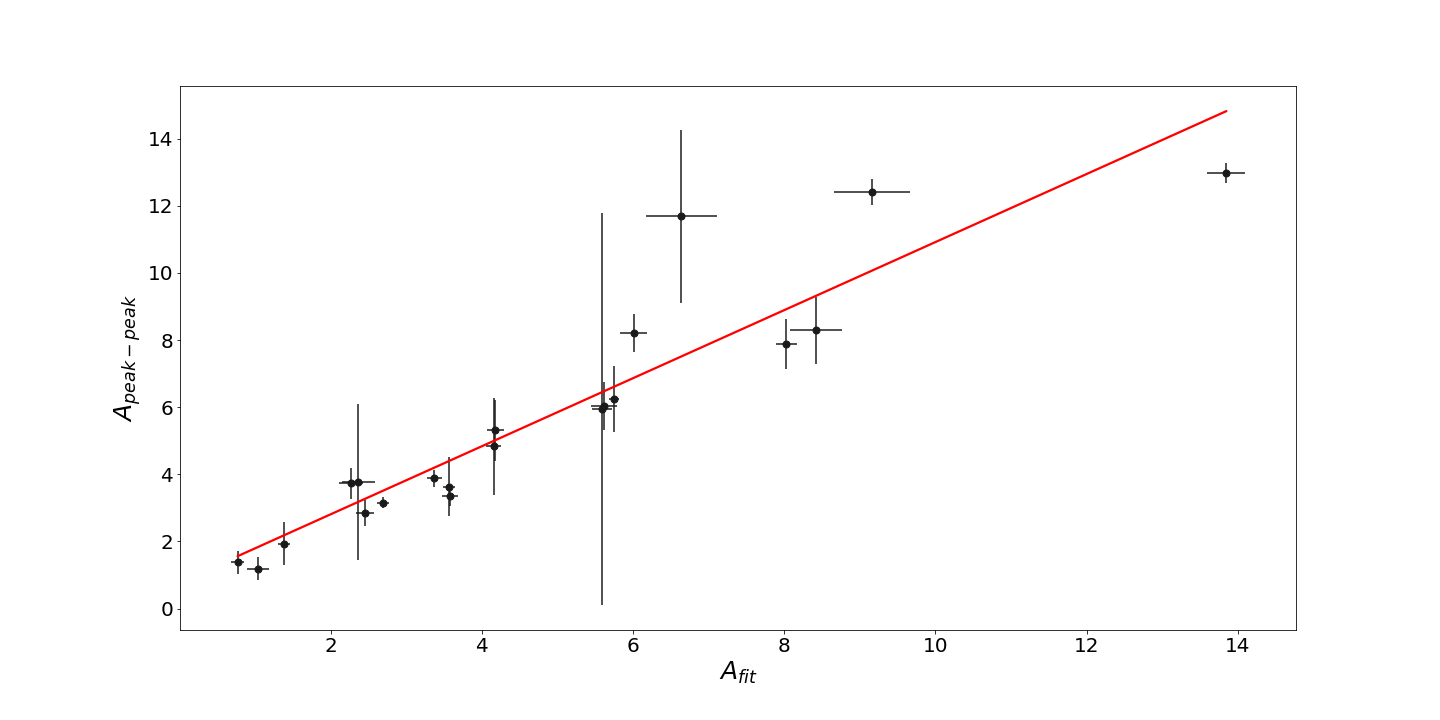}
    \caption{Comparisons of rotation amplitude values obtained from the sinusoidal best-fit function and the peak-to-peak approach. The black dots are the values of the rotation amplitude for each GCs, while the red line is the best-fit line obtained from the linear regression.}
    \label{fig:fit}
\end{figure*}

\noindent As a final analysis, a comparison between the rotation amplitude values obtained from the best-fit sine function of the observed patterns and the rotation amplitude values derived from the peak-to-peak method was made. 
As can be seen in the Table \ref{dati}, the calculated amplitudes are within the uncertainties of each other, but in some cases, there are significant differences between the two methods (e.g. 6.64 km/s versus 11.68 km/s for NGC 6388). Moreover, for {a few} GCs, the peak-to-peak approach returns {slightly} high uncertainty values. The result of this comparison is shown in Figure \ref{fig:fit}. In this figure, the values of $A_{fit}$ and $A_{peak-peak}$ (black dots) for each GCs were plotted and a linear regression was performed to find the line that best fit the data (red line). As a result, the best-fit line is:
\begin{equation}
    A_{peak-peak} = 1.0 \cdot A_{fit} + 0.8
\end{equation}

\noindent Consequently, due to the {slightly} high uncertainty values from the peak-to-peak approach {for a few GCs}, the derivation of rotation amplitude from the best-fit sine function method is the most accurate. However, given the best-fit equation, it can be concluded that the use of the two methods gives equivalent results. Indeed, both methods can be considered two good tools for determining the rotation amplitudes.

\subsection{{Comparison with literature}}
The signature of systemic rotation for these GCs is consistent with the works of \citealt{Lane10b}, \citealt{Bellazzini+12}, \citealt{Fabricius+14}; \citealt{Cordero17}; \citealt{Johnson17}, \citealt{Kamann+18}, \citealt{Bianchini18}, \citealt{Ferraro_2018}, \citealt{Lanzoni_2018}, \citealt{Sollima19} and \citealt{Szigeti+21}. Among these, of particular interest are the work of \citealt{Lane10b}, \citealt{Bellazzini+12}, \citealt{Bianchini+13}; \citealt{Lardo15}; \citealt{Kimmig+15}; \citealt{Lee_2017}; \citealt{Ferraro_2018} and \citealt{Lanzoni_2018}, because they used the same method as in this work to analyse the same clusters, and the work of \citealt{Szigeti+21} because they used data from 
APOGEE-2 survey.
Since all clusters have been studied in the literature before, it is possible to compare the results of the present homogeneous sample with these other previous studies. The latest results available in the literature were collected in Table \ref{confronto}, which contains the calculated rotation amplitude and position angles of the rotational axis. %Multiple conventions are used in the literature for angle and direction notations. %For example, \citealt{Bellazzini+12} calculated the $PA_{0}$ as the difference between 270° and the phase of best-fit sine function, namely $PA_{0}$ = 270° - $\phi$. 
Since multiple conventions are used in the literature for angle and direction notations (e.g. in the work of \citealt{Fabricius+14}; \citealt{Cordero17}; \citealt{Kamann+18} and \citealt{Sollima19}), the published results have been converted the PA 90 = East convention, anti-clockwise system. The conversion of the $PA_{0}$ values of these works can be found in the appendix \ref{conversions}. The $PA_{0}$ values measured in this work and shown in Table \ref{dati} have also been converted. To convert the values into the reference notation, a subtraction was performed between 360° and the measured $PA_{0}$ value for each cluster. \\

\noindent {As reported above, one of the most relevant works with which to compare the results is the study of \citealt{Szigeti+21}, because also in this work APOGEE data have been used. However, our study shows some differences in terms of available data. The first difference is that \citealt{Szigeti+21} used the DR14 data release of APOGEE (\citealt{Holtzman_2018}), while we used the expanded DR17 data release of APOGEE (\citealt{Schiavon23}). Two other differences are the number of GCs analysed and the stars observed for each individual cluster. Indeed, \citealt{Szigeti+21} studied a total of 10 GCs, but they were able to successfully measure the rotation speed and position angle
of the rotation axis for 9 clusters. Instead, in our work, we more than doubled the sample to a total of 23 GCs and we were able to successfully measure the rotation amplitude and the position angle of the rotation axis for 21 clusters. Moreover, the number of stars observed for each individual cluster varies from 26 to 215 in \citealt{Szigeti+21}, while it varies from 65 to 1864 in our work. Despite these differences in data, the methodology to investigate the rotation amplitude and the position angle
of the rotational axis of the clusters (namely the second method described in Section \ref{3}) is similar, thus it is possible to compare the results of the two works.}\\

\noindent {As noted from the Table \ref{confronto}, the comparison between our results and those reported in literature works shows some discrepancies. In particular, even for the best-studied case (NGC 5139 with 1864 stars), the results are not always consistent with the literature. As regards the rotation amplitude, there is a factor of $\sim$ 2 discrepancy present in almost every cluster. This can be explained by considering that the rotation amplitude is usually taken to be half of the amplitude of the sinusoidal function (see for example \citealt{Lane+11} and \citealt{Szigeti+21}). Even the position angle of the rotational axis values in this work are different from those reported in the literature. Some of these discrepancies (especially for the clusters with a lower number of stars) can be explained by different field of view used in the data (e.g. \citealt{Kamann+18} focuses on more central FoVs) or different data used (e.g. \citealt{Sollima19} derives 3D rotation amplitudes from proper motion and line-of-sight velocity simultaneously), or lack of perspective correction (e.g. \citealt{Van_de_Ven06}). {\noindent However, other effects may also be responsible for the discrepancies between this work and those of the literature (and in some cases also between one literature work and another). The first depends on the observed patterns and the definition of the best-fit sine function that fits these points. A different position of the observed patterns or different fit of the sine function can lead to discrepant results. As in the case of NGC 5139 ($\omega$ Cen) that does not show a minimum or maximum position in the best-fit sine function at low values of PA (as shown instead in the work of \citealt{Bianchini+13}), but the minimum or maximum is present to higher values of PA. This can explain the high $PA_{0}$ value obtained in this work compared to the other literature work\footnote{{A high position angle value has also been derived in \citealt{Van_de_Ven06} for $\omega$ Cen. In this work, the position angle has been defined as the angle between the observed major axis and North (measured counterclockwise through East) and determined by fitting elliptic isophotes to the smoothed Digital Sky Survey (DSS) image of the cluster, while keeping the centre fixed. In this way, the position angle is about 100°, a value discrepant to other values reported in the literature.}}. The same applies to the globular cluster NGC 6205, in which the work of \citealt{Szigeti+21} notes a minimum position at lower PA, while in this work it is to higher PA values. In both of these cases, the observed patterns, and consequently the best fits, are similar to those reported in the literature works, but with a shift which varies the $PA_{0}$ results.
A different situation is that of the globular clusters NGC 2808, NGC 3201, NGC 5904 and NGC 7078. In this case, the position of the observed patterns and the definition of the best-fit sine function are definitely divergent from those reported in the literature, causing discrepancies in $PA_{0}$ values. Furthermore, although the same method and the convention PA 90 = East anti-clockwise system were used to define the observed patterns and the best-fit sine function, different approaches have been adopted to determine $PA_{0}$ in different works. For instance, \citealt{Ferraro_2018} consider only the minimum position in the best-fit sine function to derive the $PA_{0}$ for the globular clusters NGC 5272, NGC 3201, NGC 0362, NGC 6171, NGC 6254, whereas this work takes into account both the minimum and maximum position (the same method used in \citealt{Szigeti+21}). Instead, \citealt{Lardo15} consider only the maximum position in the best-fit sine function for the clusters NGC 2808, NGC 6752, NGC 7078 and NGC 1851. Another method is the one used by \citealt{Bellazzini+12} which derives the $PA_{0}$ value from the best-fit equation using the formula $\phi = 270° - PA_{0}$ (where $\phi$ is the phase of best-fit sine function) for the globular clusters NGC 5904, NGC 2808, NGC 7078, NGC 1851, NGC 6171, NGC 6254 and NGC 6388. All these effects mentioned above may be responsible for the discrepancies in $PA_{0}$ between this work and those in the literature.}\\

\noindent {The last comparison with the literature regards the two GCs, NGC 6218 and NGC 6838, that do not show a coherent sinusoidal pattern in the figures\footnote{The reference figures are those that show the diﬀerence between the mean velocities on each side of a cluster with respect to a line passing through the cluster center with a position angle PA (measured from north to east, north = 0°, east = 90°, clockwise system), as a function of the adopted PA.} derived from the second method described in Section \ref{3} {(see Appendix \ref{GCsNoRot})}.} Consequently, there is no clear sign of rotation for these two clusters and for these clusters, it was not possible to measure $A_{rot}$ (neither with the best-fit sine function nor with the peak-to-peak approach, see section \ref{3}) and $PA_{0}$. In the literature, both clusters have been studied by \citealt{Bellazzini+12}, \citealt{Kimmig+15} and \citealt{Bianchini18} and in these works, they found a signature of rotation. As explained in \citealt{Bellazzini+12}, a work that uses the same adopted technique in this analysis, the results may also suﬀer from biases associated with the dimension of the samples for each GCs and with the radial distribution of sample stars. Moreover, the data used in \citealt{Bellazzini+12}, \citealt{Kimmig+15} and \citealt{Bianchini18} come from FLAMES-GIRAFFE sample, Hectochelle on the MMT telescope, and Gaia DR2 respectively, while the data of this work come from APOGEE-2 DR17 and Gaia (EDR3). Therefore, more observational data are required to verify the presence of systemic rotation in the GCs NGC 6218 and NGC 6838.

\subsection{{General considerations}}
\noindent {Below are two interesting considerations regarding the available data and possible effects that could influence the detection of the systemic rotation in the GCs.}\\
\noindent {An important effect that can produce a non-negligible amount of apparent
rotation is the perspective rotation. It is particularly relevant for GCs that have a
large extent on the plane of the sky. For example, for NGC 5139 ($\omega$ Cen), this effect can be extremely relevant (see for example \citealt{Merritt97} and \citealt{Van_de_Ven06}). Indeed, as reported in \citealt{Van_de_Ven06}, for the data typically extending to 20 arcmin from the $\omega$ Cen
cluster centre, the maximum amplitude of the perspective rotation for the proper motions is about 0.06 mas $yr^{-1}$ and for the line-of-sight velocity about 0.8 km $s^{-1}$. These values are a significant fraction of the observed mean velocities {shown in this literature work}, so the perspective rotation cannot be ignored. {For this reason, we calculated the perspective rotation for $\omega$ Cen, which is the GC with the largest extent on the plane of the sky in our sample of clusters, using our available data and the technique described in \citealt{Van_de_Ven06}. As $PA_{0}$ we considered the value derived in this work, namely 162°.
To derive the values of the perspective rotation for each star, we applied Eq. 6 reported in the reference literature work, adopting the canonical distance of 5 kpc and the systemic motion given in Eq. 3 from the reference work. The perspective rotation of each star is insignificant compared to its $\widetilde{V_{r}}$, so it could be ignored. Nevertheless, we corrected the observed
stellar velocities by subtracting it. As a result, the kinematic data corrected for perspective rotation return a $A_{fit}$ value of 13.69 km/s (instead of 13.85 km/s without correction) and a $A_{peak-peak}$ value of 13.32 km/s (instead of 12.98 km/s). The difference between the values with and without correction is extremely low, so the effect of perspective rotation is negligible. Since this effect is inappreciable for $\omega$ Cen and all the other clusters analyzed in this work have a Jacobi radius, which is considered as a parameter for the size of the field of view sampled, lower than that of $\omega$ Cen (see Table \ref{dati}), in this work, we did not correct the observed velocities for perspective rotation.} \\

\noindent {The strategy to use the fitting of a sine law is commonly used to detect rotation from resolved kinematics. It is a reasonably good method to estimate the PA of the rotation axis. However, for the estimation of the rotation amplitude, it presents a major drawback. Indeed, given that GCs are characterized by differential rotation curves, the value of the rotation amplitude measured with this method strongly depends on the spatial distribution of the line-of-sight velocity samples. A differential rotation profile of a GC usually peaks around 1-2 half-light radii. Therefore, if the data do not sample this region, they will systematically show a lower rotation value, and vice versa if they are concentrated around these intermediate regions. Nevertheless, as demonstrated for example in \citealt{Lanzoni_2018}, which tests the variability of their results considering different radii, the rotation amplitude and the position angle of the rotation axis are essentially constant in all of the investigated radii, as expected in the case of a coherent
global rotation of the system. In this reference work, the authors considered a set of concentric annuli around the cluster center of NGC 5904, avoiding the innermost region, where the statistic is poor, and the outermost region, where the sampling is scant and nonsymmetric. 
As a result of this test (see Table 2 and Figure 7 of the reference paper), the values of $A_{rot}$ and $PA_{0}$ are very similar between the annulus and the other. For this reason, since we consider the case of a coherent global rotation for each GC,
in this work we did not carry out a test on the variability of the results at different radii.}\\

\begin{table*}
\centering
\vspace{.1cm}
\caption{Comparison with literature values. The first sub-column represents the rotation amplitude in km/s and the second is the position angle of the rotational axis in degrees. The last row contains the values of this analysis, which represent the values of the observed rotation. Since different conventions of $PA_{0}$ were followed, the published result have been converted to PA 90 = East, anti-clockwise system.}    

\begin{center}
\tiny

\begin{tabular}{lccccc c ccccc c cccc c cccc }

 \hline
 
  Reference & \multicolumn{2}{c}{NGC 5139} &&& \multicolumn{2}{c}{NGC 6715} &&& \multicolumn{2}{c}{NGC 0104} &&& \multicolumn{2}{c}{NGC 5272}\\
\cline{2-3} \cline{6-7} \cline{10-11} \cline{14-15}   
     & $A_{rot}$   & $PA_{0}$  &&&  $A_{rot}$   & $PA_{0}$ &&& $A_{rot}$   & $PA_{0}$&&&$A_{rot}$   & $PA_{0}$\\
     \hline
\citealt{Lane10b}&-&-&&&-&-&&&2.2±0.2&\\
\citealt{Bellazzini+12}&6.0±1.0&-&&&2.0±0.5&- &&& 4.4±0.4&-&&&-&-\\
\citealt{Bianchini+13}&6.79&12±1&&&-&-&&&4.00&136±1&&&-&- \\
\citealt{Fabricius+14}&-&-&&&-&-&&&-&-&&&-&192.2±11.8\\
\citealt{Kimmig+15}&-&-&&&1.6±2.9&-&&&4.0±0.3&-&&&0.6±1.0&- \\
\citealt{Ferraro_2018}&-&-&&&-&-&&&-&-&&&1.0&151\\
\citealt{Kamann+18}&-&9.9±4.3&&&-&-&&&-&{134.1±3.6}&&&-&-\\
\citealt{Sollima19}&4.27±0.52&9.8±7.6&&&0.57±1.11&-&&&5.00±0.32&135.7±4.6&&&1.75±0.42&-\\
\citealt{Szigeti+21} &-&-&&&-&-&&&-&-&&&1.19±0.3&164±15\\
\hline
This work&13.85±0.25&162&&&6.01±0.18&252&&&9.16±0.50&216&&&2.46±0.12&162\\
\hline
     
\label{tab:star}

\end{tabular}

\end{center}

%\vspace{0.005mm}

\begin{center}
\tiny

\begin{tabular}{lccccc c ccccc c cccc c cccc }

 \hline
 
  Reference & \multicolumn{2}{c}{NGC 6656} &&& \multicolumn{2}{c}{NGC 3201} &&& \multicolumn{2}{c}{NGC 5904} &&& \multicolumn{2}{c}{NGC 6121}\\
\cline{2-3} \cline{6-7} \cline{10-11} \cline{14-15}   
     & $A_{rot}$   & $PA_{0}$  &&&  $A_{rot}$   & $PA_{0}$ &&& $A_{rot}$   & $PA_{0}$&&&$A_{rot}$   & $PA_{0}$\\
     \hline
\citealt{Lane10b} &1.5±0.4&-&&&-&-&&&-&-&&&0.9±0.1&70$-$250\\
\citealt{Bellazzini+12}& 1.5±0.4&-&&& 1.2±0.3&-   &&&2.6±0.5&157&&&1.8±0.2&-   \\
\citealt{Fabricius+14}&-&-&&&-&-&&&-&148.5±5.6&&&-&-\\
\citealt{Kimmig+15}&2.5±2.3&-&&&-&-&&&2.1±0.7&-&&&1.3±0.5&- \\
\citealt{Lee_2017}&-&-&&&-&-&&&3.36±0.7&128&&&-&- \\
\citealt{Ferraro_2018}&-&-&&&1.3&215&&&-&-&&&-&-\\
\citealt{Lanzoni_2018}&-&-&&&-&-&&&4.0&145&&&-&-\\
\citealt{Kamann+18}&-&{280.9±23.9}&&&-&{244.3±188.9}&&&-&{305.7±20.3}&&&-&{214.5±9.4}\\
\citealt{Sollima19}&3.38±0.71&107.2±9.2&&&0.80±0.41&-&&&4.11±0.42&138.4±6.0&&&0.22±0.17&-\\
\citealt{Szigeti+21}&-&-&&&-&-&&&3.45±0.4&148±6&&&-&- \\

\hline
This work&5.75±0.06&252&&&3.37±0.10&324&&&8.03±0.14&198&&&2.70±0.08 &234 \\
\hline
     
\label{tab:star}

\end{tabular}

\end{center}
%\vspace{0.005mm}

\begin{center}
\tiny

\begin{tabular}{lccccc c ccccc c cccc c cccc }

 \hline
 
  Reference & \multicolumn{2}{c}{NGC 2808} &&& \multicolumn{2}{c}{NGC 6205} &&& \multicolumn{2}{c}{NGC 6752} &&& \multicolumn{2}{c}{NGC 6397}\\
\cline{2-3} \cline{6-7} \cline{10-11} \cline{14-15}   
     & $A_{rot}$   & $PA_{0}$  &&&  $A_{rot}$   & $PA_{0}$ &&& $A_{rot}$   & $PA_{0}$&&&$A_{rot}$   & $PA_{0}$\\
     \hline
\citealt{Lane10b}&-&-&&&-&-&&&$\leq 0.2$&-&&&-&- \\
\citealt{Bellazzini+12} &3.3±0.5&152&&&  -&-&&& 0.0±0.0&-&&&0.2±0.5&-    \\
\citealt{Fabricius+14}&-&-&&&-&196.5±7.8&&&-&-&&&-&-\\
\citealt{Lardo15}&4.72±0.2&270&&&-&-&&&0.67±0.2&200&&&-&- \\
\citealt{Kimmig+15}&3.1±3.8&-&&&-&-&&&0.3±0.5&-&&&-&- \\
\citealt{Cordero17}&-&-&&&2.7±0.9&14±19&&&-&-&&&-&- \\
\citealt{Kamann+18}&-&{313.0±2.4}&&&-&-&&&-&{139.1±41.8}&&&-&-\\
\citealt{Sollima19}&2.25±0.56&143.9±8.4&&&1.53±0.61&14.5±14.2&&&0.91±0.34&-&&&0.48±0.17&171.4±15.6\\
\citealt{Szigeti+21}&-&-&&&2.38±0.4&26±9&&&-&-&&&-&- \\
\hline
This work&8.42±0.34&198&&&5.62±0.17&306&&&1.04±0.15&162&&&1.39±0.08&198\\
\hline
     
\label{tab:star}

\end{tabular}

\end{center}

%\vspace{0.005mm}

\begin{center}
\tiny

\begin{tabular}{lccccc c ccccc c cccc c cccc }

 \hline
 
  Reference & \multicolumn{2}{c}{NGC 7078} &&& \multicolumn{2}{c}{NGC 0362} &&& \multicolumn{2}{c}{NGC 1851} &&& \multicolumn{2}{c}{NGC 6171}\\
\cline{2-3} \cline{6-7} \cline{10-11} \cline{14-15}   
     & $A_{rot}$   & $PA_{0}$  &&&  $A_{rot}$   & $PA_{0}$ &&& $A_{rot}$   & $PA_{0}$&&&$A_{rot}$   & $PA_{0}$\\
     \hline
\citealt{Bellazzini+12}&3.8±0.5&290&&&-&-&&& 1.6±0.5&252&&&2.9±1.0&264      \\
\citealt{Bianchini+13}&2.84&106±1&&&-&-&&&-&-&&&-&- \\
\citealt{Lardo15}&3.63±0.1&120&&&-&-&&&1.65±0.5&50&&&-&- \\
\citealt{Kimmig+15}&2.5±0.8&-&&&1.6±1.4&-&&&-&-&&&-&- \\
\citealt{Ferraro_2018}&-&-&&&1.1&260&&&-&-&&&1.2&167\\
\citealt{Kamann+18}&-&{150.9±10.4}&&&-&{318.0±117.6}&&&-&3.2±3.5&&&-&-\\
\citealt{Sollima19}&3.29±0.51&127.4±28.8&&&0.51±0.56&-&&&0.45±0.42&-&&&0.70±0.46&-\\
\citealt{Szigeti+21}&2.38±0.4&120±11&&&-&-&&&-&-&&&0.72±0.3&168±30 \\
\hline
This work&4.16±0.10&234&&&3.57±0.08&234&&&5.59±0.13&180&&&0.77±0.09&234\\
\hline
     
\label{tab:star}

\end{tabular}

\end{center}
%\vspace{0.005mm}

\begin{center}
\tiny

\begin{tabular}{lccccc c ccccc c cccc c cccc }

 \hline
 
  Reference & \multicolumn{2}{c}{NGC 6254} &&& \multicolumn{2}{c}{NGC 6273} &&& \multicolumn{2}{c}{NGC 6341} &&& \multicolumn{2}{c}{NGC 6809}\\
\cline{2-3} \cline{6-7} \cline{10-11} \cline{14-15}   
     & $A_{rot}$   & $PA_{0}$  &&&  $A_{rot}$   & $PA_{0}$ &&& $A_{rot}$   & $PA_{0}$&&&$A_{rot}$   & $PA_{0}$\\
     \hline
\citealt{Lane10b}&-&-&&&-&-&&&-&-&&&0.25±0.09&-\\
\citealt{Bellazzini+12} &0.4±0.5& 95&&&-&-&&&-&-&&&0.5±0.2& -     \\
\citealt{Fabricius+14}&-&153.5±14.7&&&-&-&&&-&98.9±12.0&&&-&-\\
\citealt{Kimmig+15}&-&-&&&-&-&&&1.8±0.8&-&&&0.4±0.2&- \\
\citealt{Johnson17}&-&-&&&3.83±0.12&126±2&&&-&-&&&-&-\\
\citealt{Ferraro_2018}&1.4&315&&&-&-&&&-&-&&&-&-\\
\citealt{Kamann+18}&-&{142.8±17.9}&&&-&-&&&-&-&&&-&-\\
\citealt{Sollima19}&0.26±0.56&-&&&4.19±1.12&123.1±13.2&&&1.46±0.61&-&&&0.88±0.38&-\\
\citealt{Szigeti+21}&-&-&&&-&-&&&2.06±0.6&154±14&&&-&- \\
\hline
This work&2.27±0.16&198&&&2.37±0.22&270&&&4.18±0.11&216&&&3.58±0.10&252\\
\hline
     
\label{tab:star}

\end{tabular}

\end{center} 

%\vspace{0.005mm}
\begin{center}
\tiny

\begin{tabular}{lccccc c ccccc c cccc c cccc }

 \hline
 
  Reference & \multicolumn{2}{c}{NGC 6388} \\
\cline{2-3}    
     & $A_{rot}$   & $PA_{0}$ \\
     \hline
\citealt{Bellazzini+12} &3.9±1.0&117        \\
\citealt{Kamann+18}&-&{318.2±16.8}\\
\citealt{Sollima19}&1.51±0.65&-\\
\hline
This work&6.64±0.47&270\\
\hline
     
\label{tab:star}

\end{tabular}

\end{center} 

\label{confronto}

\end{table*}

\section{Conclusions}\label{5}
In this work, 23 globular clusters have been studied to verify the presence or not of systemic rotation. Below are the main conclusions of this analysis:
\begin{itemize}
    \item[$\bullet$] Using the data and methods described in this analysis, 21 GCs show a signature of systemic rotation. For these clusters, the rotation amplitude and the PA of the rotation axis were calculated. As a result, the clusters cover a remarkable range of rotational amplitudes, from $\sim$ 0.77 km/s to $\sim$ 13.85 km/s. In particular, the clusters that exhibit highest rotation are {NGC 5904, NGC 5139 and NGC 0104.} \\
    \item[$\bullet$] Two different methods have been used to calculate the rotation amplitude: the best-fit sine function and the peak-to-peak methods. A comparison between the two methods shows they give similar results, within the uncertainties.\\
    \item[$\bullet$] The comparison with the literature shows that in general our measurements are consistent with previously published rotation values.
    
\end{itemize}

\noindent In conclusion, most if not all of the present-day GCs show measurable rotation. Processes like interactions can induce rotation in some individual cases, but it is very difficult to spin up individually the whole population of such massive objects ($ > 10^4 M_{\odot}$).
Therefore we surmise that this trait was probably imprinted initially, during the epoch of their process of formation {(see for example \citealt{Kamann+18} and \citealt{Bianchini18})}. With the advent of large spectroscopic surveys like 4MOST (\citealt{Chiappini+19}) and WEAVE (\citealt{Jin+23}), it would be possible to explore in more detail the rotational properties of larger homogeneous GCs samples.

\begin{acknowledgements}\\
  %    Part of this work was supported by the German
   %   \emph{Deut\-sche For\-schungs\-ge\-mein\-schaft, DFG\/} project
    %  number Ts~17/2--1.
{I.P. acknowledges support from ANID BECAS/DOCTORADO NACIONAL 21230761.}\\
J.G.F-T acknowledges support provided by Agencia Nacional de Investigaci\'on y Desarrollo de Chile (ANID) under the Proyecto Fondecyt Iniciaci\'on 2022 Agreement No. 11220340, and from ANID under the Concurso de Fomento a la Vinculaci\'on Internacional para Instituciones de Investigaci\'on Regionales (Modalidad corta duraci\'on) Agreement No. FOVI210020, and from the Joint Committee ESO-Government of Chile 2021 under the Agreement No. ORP 023/2021, and from Becas Santander Movilidad Internacional Profesores 2022, Banco Santander Chile, and from Vicerrector\'ia de Investigaci\'on y Desarrollo Tecnol\'ogico (VRIDT) at Universidad Cat\'olica del Norte under resoluci\'on No. 061/2022-VRIDT-UCN.\\

D.M. gratefully acknowledges support from the ANID BASAL projects ACE210002 and FB210003, from Fondecyt Project No. 1220724, and from CNPq Project 350104/2022-0.

\end{acknowledgements}

% WARNING
%-------------------------------------------------------------------
% Please note that we have included the references to the file aa.dem in
% order to compile it, but we ask you to:
%
% - use BibTeX with the regular commands:
%   \bibliographystyle{aa} % style aa.bst
%   \bibliography{Yourfile} % your references Yourfile.bib
%
% - join the .bib files when you upload your source files
%-------------------------------------------------------------------

\clearpage
\bibliographystyle{aasjournal}
\bibliography{bibliography}

\begin{appendix}%First appendix
\onecolumn
\section{$PA_{0}$ conversions}\label{conversions}
This Appendix includes the details of the conversions of the position angle of the rotation axis values reported in the works of \citealt{Fabricius+14}, \citealt{Cordero17}, \citealt{Kamann+18}, and \citealt{Sollima19} into the north, PA = 0°, to east, PA = 90°, anti-clockwise system (traditional reference system). In each case, $PA_{0}$ is the final position
angle of the rotation axis used in this paper, after conversion into the north, PA = 0°, to east, PA = 90°, anti-clockwise system, and $PA_{0,*}$ is the position angle of the rotation axis reported in the studies listed above.\\

\begin{itemize}
    \item[$\bullet$] In \citealt{Fabricius+14}, the kinematic position angle has been obtained from the velocity field. Since the values of $PA_{0,*}$ of GCs in common with this work cover between 0° and 180°, the value of $PA_{0}$ in the traditional reference system is equal to the sum of the $PA_{0,*}$ and 90°.\\
    \item[$\bullet$] In the case of \citealt{Cordero17}, the value of the $PA_{0}$ is that converted and reported in \citealt{Szigeti+21}.\\

    \item[$\bullet$] In \citealt{Kamann+18}, the values of $PA_{0}$ are written in the PA = 0°, to east, PA = 90°, anti-clockwise system. However, the range of $PA_{0,*}$ values does not cover from 0° to 360° (where all values are positive). Indeed, in this work, the ranges of $PA_{0,*}$ values are between [0°, 180°], [-180°, -0°]. Therefore, to convert the reported values, the following algorithm is used:
    \begin{itemize}
        \item {If 0°<$PA_{0,*}$<180°, then the value of $PA_{0}$ in the traditional reference system is equal to $PA_{0,*}$;}
        %\item If -90°<$PA_{0,*}$<-180° (namely 180°<$PA_{0}$<270° in the traditional reference system), then $PA_{0}$ = (|$PA_{0,*}$| - 90°) + 180°;
        \item {If 0°<$PA_{0,*}$<-180° (namely 180°<$PA_{0}$<360° in the traditional reference system), then $PA_{0}$ = 360° - |$PA_{0,*}$|}\\
        %\item If 90°<$PA_{0,*}$<180° (namely 270°<$PA_{0}$<360° in the traditional reference system), then $PA_{0}$ = ($PA_{0,*}$ - 90°) + 270°\\
        
    \end{itemize}

    \item[$\bullet$] \citealt{Sollima19} used the convention from north, PA = 0°, to west, PA = 90°, anti-clockwise system. Therefore, in this case the following algorithm is used:
    \begin{itemize}
        \item If 0°<$PA_{0,*}$< 180°, then the value of $PA_{0}$ in the traditional reference system is 180°- $PA_{0,*}$
         \item Otherwise, if $PA_{0,*}$>180°, then the value of $PA_{0}$ in the traditional reference system is 360°- $PA_{0,*}$ 
      \end{itemize}
    \end{itemize}

\vspace{5mm}

\section{GCs without signature of systemic rotation}\label{GCsNoRot}
{This Appendix reports the difference between the mean radial velocities located on each side of the cluster with respect to a line passing through the center as a
function of PA (measured from north, PA = 0°, to east, PA = 90°, clockwise system) as a function of the adopted PA, for the globular clusters NGC 6218 and NGC 6838. These figures do not show coherent sinusoidal behavior of the observed patterns (blue dots), thus these clusters do not show a signature of systemic rotation.}

%\begin{multicol}
\begin{figure*}[h]

  \centering
  \includegraphics[width=.49\textwidth]{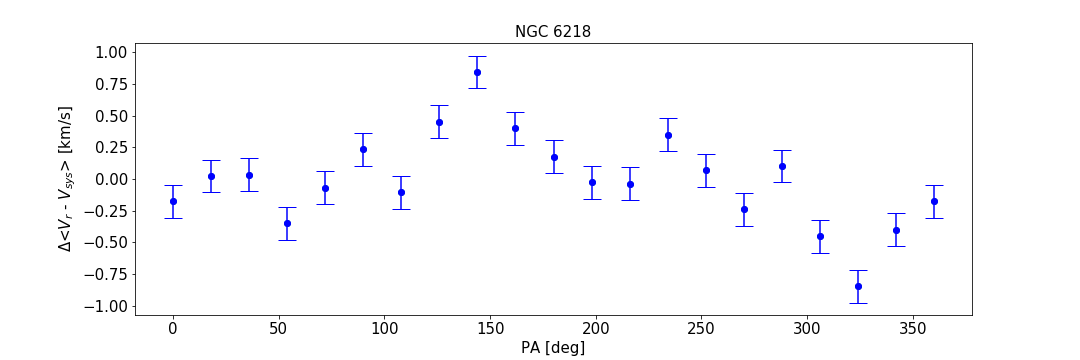}
  \hspace{0.01cm}
  \includegraphics[width=.49\textwidth]{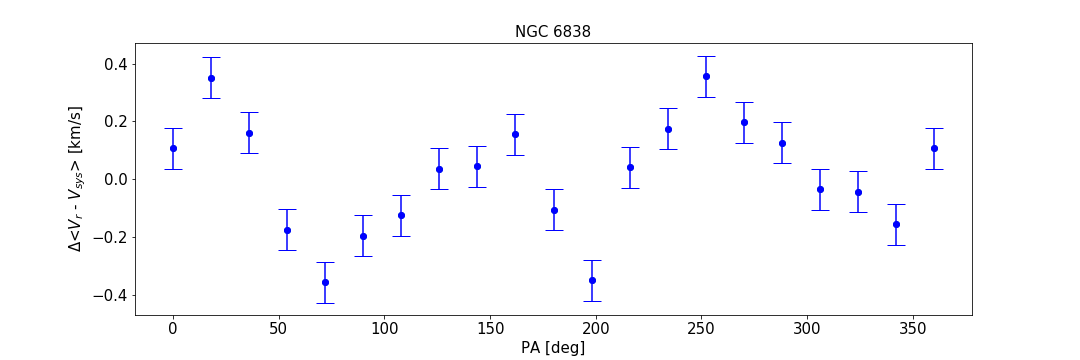}\\

  \vspace{0.5cm}
  \caption{Left: NGC 6218. Right: NGC 6838.}
\end{figure*}

%\end{multicol}

\section{Test to assess step value}\label{app_test}
\noindent {This Appendix explores the choice of using a step of 18° in the second and third methods described in Section \ref{3} to calculate the  $A_{rot}$ and $PA_{0}$. Indeed, especially the peak-to-peak approach is very likely strongly affected by statistical scatter due to the determination of the peak and by the steps in degree used for the construction of the curves. Therefore, the value of 18°, used for all clusters, might seem rather large, especially for those GCs, such as NGC 5139, where a large number of line-of-sight velocities are available. For these reasons, we investigate the dependence of our results on the step chosen to demonstrate the robustness of our methods.\\
This appendix shows the results obtained for the NGC 5139 cluster, which is the GC with the highest number of available stars, using the same methods adopted in our analysis and described in Section \ref{3}, but with a step of 9°. The signature of systemic rotation using a step of 9° for this cluster is shown in Figure \ref{fig:ngc5139_test}, while the $A_{rot}$ and $PA_{0}$ obtained using this step and the comparison with the results obtained with a step of 18° is shown in the Table \ref{tab:confronto_step}. \\
As a result, the use of these two different steps gives equivalent values of $A_{rot}$ and $PA_{0}$. Therefore, the use of a step of 18° for all the clusters is a reasonable choice. Indeed, it allows us to consider a significant number of stars for each range in the case of clusters with a low number of available stars. Whereas, in the case of GCs with a high number of available stars, the use of this step does not show a significant difference in the results. }

\begin{figure*}[ht]
    \centering
    \includegraphics[width=0.8\textwidth]{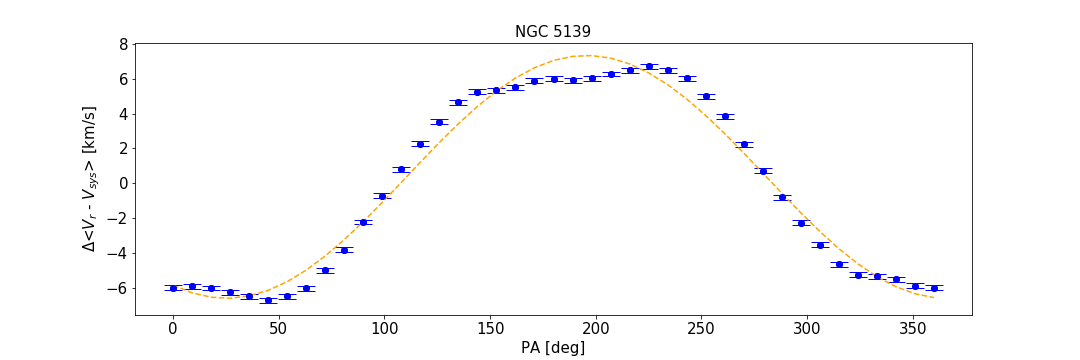}
    \caption{{Difference between the mean radial velocities located on each side of the cluster with respect to a line passing through the center as a function of PA (measured from north, PA = 0°, to east, PA = 90°, with a step of 9°, clockwise system) as a function of the adopted PA, for the globular cluster NGC 5139. The orange dashed line is the sine function that best fits the observed patterns (blue dots).}}
    \label{fig:ngc5139_test}
\end{figure*}

\begin{table*}
\centering
\vspace{.1cm}
\caption{{Comparison of the results obtained using a step of 18° and a step of 9° for the globular cluster NGC 5139.}}
\label{tab:confronto_step}
\begin{tabular}{lccr}

\hline
\hline
\multicolumn{3}{c}{{NGC 5139}}\\
\hline
&\multicolumn{1}{c}{Step of 18°} & \multicolumn{1}{c}{Step of 9°}\\
\hline

\hline
$A_{fit}$ & 13.85 ± 0.25 & 13.94 ± 0.17 \\
$A_{peak-peak}$ & 12.98 ± 0.29& 13.43 ± 0.20 \\
$PA_{0}$& 162 & 162\\
\hline

\end{tabular}
\end{table*}

\end{appendix}

\end{document}